\documentclass[a4j,11pt,onecolumn,oneside,notitlepage,final]{article}

\usepackage{amsmath}
\usepackage{amsfonts}
\usepackage{amssymb}
\usepackage{latexsym}
\usepackage{color}
\usepackage{graphicx}
\usepackage{cite}
\usepackage{subfigure}
\usepackage{ascmac}
\usepackage{authblk}

\def\be{\begin{equation}}
\def\ee{\end{equation}}
\def\ben{\begin{eqnarray}}
\def\een{\end{eqnarray}}
\def\da{{\dot a}}

\begin{document}
	\begin{center}
		\hfill RESCEU-10/19 \\
		\vskip .5in
		
		{\Large \bf
			Formation threshold of rotating primordial black holes
		}
		\vskip .45in

{
Minxi He$^{a,b}$,
Teruaki Suyama$^{c}$
}

{\em
$^a$
   Department of Physics, Graduate School of Science, The University of Tokyo, Tokyo 113-0033, Japan
}\\
{\em
$^b$
   Research Center for the Early Universe (RESCEU), Graduate School of Science, The University of Tokyo, Tokyo 113-0033, Japan
}\\
{\em
$^c$
   Department of Physics, Tokyo Institute of Technology,
	2-12-1 Ookayama, Meguro-ku, Tokyo 152-8551, Japan
}\\

\abstract{
	Within the framework that primordial black holes are formed by the direct gravitational collapse of  
	large primordial density perturbations in the radiation dominated stage,
	we derive the threshold of the density contrast for the formation of rotating primordial black holes
	based on the simple Jeans criterion. 
	It is found that the threshold value increases in proportion to the	square of the angular momentum. 
	We then apply the recently refined analysis on the formation threshold for non-rotating
	black holes to the case of rotating black holes, and contrast the derived threshold with the former.
	Caveats and effects ignored in our analysis are also presented, which suggests that the uncertainties
	of our result can be addressed only by means of numerical relativity. 
}

\end{center}
\vskip .4in

\tableofcontents 

\newpage

\section{Introduction}
Detections of gravitational waves by LIGO/Virgo have revealed that many 
BH binaries, which merge within the Hubble time, populate in the Universe \cite{LIGOScientific:2018mvr}.
Among several scenarios that can explain the LIGO events, one possibility
is that all of (or some of) the LIGO events were caused by the mergers of primordial black holes (PBHs) 
(see e.g. \cite{Sasaki:2018dmp}).
Contrary to the BHs of astrophysical origin, PBHs are directly produced in the very early Universe
by the gravitational collapse of  large primordial density perturbations generated during inflation.
Some fraction of the PBHs afterwards form binaries when the Universe is still in the radiation dominated epoch,
and some of the PBH binaries merge within the age of the Universe \cite{Nakamura:1997sm}.
Quantities related to PBHs such as mass and abundance contain information of inflation at different 
stage from the one probed by the CMB and the large-scale structures.
Thus, proof or disproof of the PBH hypothesis provides a useful constraint on inflation and physics of the
extremely early Universe independently of the larger scale observations. 
Given that detection of a huge number of the merger events in the next decade is promising, 
it will become feasible to test the PBH hypothesis by the gravitational-wave observations.
To this end, it is necessary to give theoretical predictions on observables and to clarify their
usefulness and powerfulness.
Several proposals have already been made, which include the cosmic evolution of the merger rate \cite{Nakamura:2016hna},
stochastic gravitational-wave background \cite{Ioka:1998gf, Mandic:2016lcn, Cholis:2016xvo, Wang:2016ana, Cai:2018dig, Cai:2019jah}, 
and the distribution of the merger rate in the two-dimensional
mass plane \cite{Kocsis:2017yty, Liu:2018ess}.

In this paper, we focus on the spin of the PBHs formed in the radiation dominated era
\footnote{Spin of the PBHs formed in the matter dominated era was studied in \cite{Harada:2017fjm}.}.
The gravitational waveform from a BH binary carries information of the spin of the individual BHs
in the binary.
After a sufficient number of the merger events are accumulated, it will become possible to
discuss the statistical distribution of the BH spin, which is expected to shed light on the 
origin of the LIGO events, e.g. \cite{Fernandez:2019kyb}.
Roughly speaking, a PBH forms if the density contrast of the overdense region 
exceeds a threshold value $\delta_{\rm th}$ \cite{Carr:1975qj} \footnote{More precisely, 
$\delta_{\rm th}$ depends on the density profile of the overdense region \cite{Niemeyer:1999ak, Shibata:1999zs, Polnarev:2006aa, Nakama:2013ica}.}
\footnote{Another convenient quantity for the criterion of the PBH formation is the compaction function
\cite{Shibata:1999zs, Musco:2018rwt, Germani:2018jgr}.}.
When the overdense region is spherically symmetric, analytic calculation shows $\delta_{\rm th}=w$ \cite{Carr:1975qj},
where $w =P/\rho$ is the equation of state parameter of the fluid dominating the Universe at the time
of the PBH formation (we are primarily interested in the case $w=1/3$
\footnote{For the case when a simple fluid description of matter is invalid, see e.g. \cite{Hidalgo:2017dfp}.}).
The refined analytic calculation, which reproduces the numerical 
results \cite{Musco:2012au, Musco:2018rwt} 
more accurately, was obtained in \cite{Harada:2013epa}.
Since the angular momentum effectively reduces the gravitational force by the centrifugal force, 
it is natural to expect that $\delta_{\rm th}$ for a rotating BH is higher than that for a non-rotating one.
In other words, $\delta_{\rm th}$ would be a function of the angular momentum $J$ of the overdense region.
In Ref.~\cite{Chiba:2017rvs}, $\delta_{\rm th}(J)$ was obtained for the critical collapse regime, but
it is not clear how many uncertainties remain in the derivation of this threshold because they focus on 
asymptotically flat spacetime instead of an expanding universe as well as the critical collapse \cite{Niemeyer:1997mt,Yokoyama:1998xd}.
The purpose of this paper is to derive an explicit form of $\delta_{\rm th}(J)$ by generalizing the arguments
which, based on the Jeans criterion, were developed in \cite{Carr:1975qj, Harada:2013epa} for the case of the non-rotating PBHs.
In the next section, we derive the threshold value by following the simple argument in \cite{Carr:1975qj}.
We then proceed to the analysis based on the refined argument developed in \cite{Harada:2013epa}
\footnote{The refined criterion was also
adopted to study the collapse of an oscillating free massive scalar field in \cite{Hidalgo:2017dfp}.}.

Before closing this section, let us briefly mention how the result derived in this paper is embedded
in the whole program of evaluating the spin distribution of PBHs.
Ignoring growth of the BH mass after the PBH formation and the critical phenomena \cite{Niemeyer:1997mt}
and assuming that the angular momentum is conserved during the collapse,
the Press-Schechter formalism tells us that the spin distribution of PBHs with its mass $M$ 
at cosmic time $t$ is formally written as 
\be
W(J,t) =\int dJ'~Q(J,J',t) \int_{\delta_{\rm th} (J')}~P(\delta_M, J') d \delta_M. \label{intro1}
\ee
Here $P(\delta_M,J)$ is the probability distribution of the density contrast $\delta_M$ and
the angular momentum $J$ of an overdense region that collapses to PBH if $\delta_M \ge \delta_{\rm th}$.
Functional form of $P(\delta_M,J)$ is determined once the underlying inflation model is assumed,
and $Q(J,J',t)$ represents the evolution of the PBH spin caused by the surrounding material 
from its initial value $J'$ to $J$ at time $t$.
All these quantities must be determined in order to derive $W(J,t)$.
In \cite{Chiba:2017rvs, Mirbabayi:2019uph, DeLuca:2019buf}, analyses related to $P(\delta_M,J)$ and $Q(J,J',t)$ have been performed.
In this paper, we address $\delta_{\rm th} (J)$.

\section{Simple estimation of the threshold of the PBH formation}
\label{simple}
It was argued in \cite{Carr:1975qj} that the threshold $\delta_{\rm th}$ of the density contrast for the PBH formation is obtained by requiring that
the size of the overdense region when it stops its expansion and turns into contraction is larger than the Jeans length.
The application of this criterion to spherically symmetric overdensity $\delta_{\rm th}$ is presented in \cite{Carr:1975qj}.
For completeness, let us first briefly describe how the threshold is derived for the spherical case quantitatively. 

A starting point is to assume that the overdense region,
whose proper size is initially super-Hubble, has uniform overdensity and to approximate the region as part of the closed FLRW Universe.
Thus, the evolution of the overdense region evolves according to the Friedmann equation given by
\be
H^2=\frac{8\pi G}{3} {\bar \rho}(1+\delta )-\frac{1}{a^2},
\ee
where ${\bar \rho}$ is the background energy density.
Because of the curvature term, the region turns from the expansion phase to the contraction phase.
Let subscript ``${\rm max}$" and ``${\rm hc}$" represent quantities at the time of the maximum expansion and the 
time when the proper size of the overdense region becomes equal to the Hubble horizon (i.e. horizon reentry), respectively.
Then, the proper size of the overdense region at the maximum-expansion time is $(a_{\rm max}/a_{\rm hc}) H_{\rm hc}^{-1}$
and the Jeans length evaluated at this time is $R_J=c_s / \sqrt{G {\bar \rho}_{\rm max} (1+\delta_{\rm max})} \simeq c_s a_{\rm max}$,
where $c_s (=1/\sqrt{3})$ is the sound speed of radiation.
Since the standard Jeans length is derived for the perturbations defined on the static and uniform fluid,
there will be ${\cal O}(1)$ ambiguity in the Jeans length given above.
This point will be addressed in the following section. 
In the present section, we take $R_J=c_s a_{\rm max}$ for definiteness.
Thus, the Jeans criterion imposed at the maximum-expansion time is written as
\be
\frac{a_{\rm max}}{a_{\rm hc}} \frac{1}{H_{\rm hc}} > c_s a_{\rm max}. \label{Jeans-criterion-no-rot}
\ee
PBH would be formed when this condition is met.
In terms of the density contrast at the horizon crossing \footnote{
To be more precise, the density contrast is defined on the uniform Hubble slicing.},
we have a relation between $a_{\rm max}$ and $a_{\rm hc}$ given by
\be
\frac{a_{\rm max}}{a_{\rm hc}} = \sqrt{\frac{1+ \delta_{\rm hc}}{ \delta_{\rm hc}}}. \label{rel-1}
\ee
Using the constancy of $\rho (1+\delta) a^4$, we also have
\be
H_{\rm hc} a_{\rm max}=\frac{\sqrt{1+\delta_{\rm hc}}}{\delta_{\rm hc}}. \label{rel-2}
\ee
Plugging Eqs.~(\ref{rel-1}) and (\ref{rel-2}) into Eq.~(\ref{Jeans-criterion-no-rot}), 
the Jeans criterion boils down to \cite{Carr:1975qj}
\be
\delta_{\rm hc} > \delta_{\rm th}=c_s^2. \label{carr-threshold}
\ee

Let us now add a small rotation to the spherically symmetric overdense region 
and consider how the rotation changes the threshold value for the PBH formation given by Eq.~(\ref{carr-threshold}).
The original argument that the size of the overdense region at the maximum expansion needs to be greater
than the Jeans length to form a BH should still hold in this case.
What needs to be modified is the Jeans length, namely the right-hand side of Eq.~(\ref{Jeans-criterion-no-rot}).
The dispersion relation for the perturbations on the rotating system with a constant 
angular velocity $\Omega$ depends on the angle between the propagation direction of the perturbation
and the rotation axis \cite{Chandra}.
Here, we adopt a criterion that the BH is formed when the Jeans stability is violated for any propagation direction 
of the perturbations, which will provide a conservative upper limit on the threshold. 
This amounts to using the dispersion relation given by \cite{Chandra}
\be
\omega^2=c_s^2 k^2+4\Omega^2-4\pi G \rho.
\ee
From this equation, we find that the Jeans length in the present case should be modified as
\be
R_J=\frac{c_s}{\sqrt{G\rho-\frac{\Omega^2}{\pi}}} \simeq \frac{c_s}{\sqrt{G\rho}} 
\left( 1+\frac{\Omega^2}{2\pi G\rho} \right). \label{Jeans-rot}
\ee
In the last equation, we have picked up only the leading term of the angular velocity, 
which is the approximation we will make throughout this paper.
This equation shows that rotation enhances the Jeans length and its effect in the slow-rotation limit is
second order in the rotational velocity. 
This is consistent with our intuition that rotation impedes gravitational contraction.

For a spherical overdense region with uniform density and uniform angular velocity, the
angular momentum is written in terms of mass $M$, the angular velocity $\Omega$, and radius $R$ of the overdense region as
\be
J=\frac{2(1+c_s^2)MR^2 \Omega}{5}.
\ee
Replacing $\Omega$ and $\rho$ by $J$ and $M$, the Jeans length (\ref{Jeans-rot}) now reads
\be
R_J=\frac{c_s}{\sqrt{G\rho}} \left( 1+\frac{25J^2}{6{(1+c_s^2)}^2 GM^3 R} \right).
\ee
Applying this Jeans length to the overdense region at the maximum-expansion time,
the Jeans criterion becomes
\be
\frac{a_{\rm max}}{a_{\rm hc}} \frac{1}{H_{\rm hc}} > c_s a_{\rm max} \left( 1+\frac{25J_{\rm max}^2}{6{(1+c_s^2)}^2 GM_{\rm max}^3} 
\frac{a_{\rm hc} H_{\rm hc}}{a_{\rm max}} \right)
. \label{Jeans-criterion-with-rot}
\ee
Let us assume that the mass and the angular momentum are both conserved during the gravitational collapse 
after the maximum expansion.
Then, mass and angular momentum of the resultant BH are given by those of the overdense region at the maximum-expansion time.
Introducing the dimensionless spin parameter of the BH as $a_K=J/(GM^2)$, 
the right-hand side of the above condition becomes
\be
c_s a_{\rm max} \left( 1+\frac{25a_K^2}{12{(1+c_s^2)}^2} 
\left( \frac{a_{\rm hc}}{a_{\rm max}} \right)^2 \frac{2GM_{\rm hc}}{H^{-1}_{\rm hc}} \right).
 \label{Jeans-criterion-with-rot2}
\ee
Thus, using Eqs.~(\ref{rel-1}) and (\ref{rel-2}) as in the non-rotational case, 
Eq.~(\ref{Jeans-criterion-with-rot}) becomes
\be
\delta_{\rm hc} > c_s^2 \left( 1+\frac{25a_K^2}{6{(1+c_s^2)}^2} \frac{\delta_{\rm hc}}{1+\delta_{\rm hc}}\right), \label{Jeans-criterion-with-rot3}
\ee
where we have made the approximation $\frac{2GM_{\rm hc}}{H_{\rm hc}^{-1}} =1$.
In the slow-rotation approximation, the above inequality can be solved as
\be
\delta_{\rm hc}>\delta_{\rm th}=c_s^2 \left( 1+\frac{25c^2_s a_K^2}{6{(1+c_s^2)}^3} \right). \label{threshold-spin}
\ee 
This is the formation threshold for the rotating PBH, which has been obtained by directly applying 
the original argument in \cite{Carr:1975qj} to the rotating case.
There are a few remarks at this point.
Firstly, as it should be, the above threshold coincides with the known result (i.e. Eq.~(\ref{carr-threshold})) 
in the non-rotation limit ($a_K \to 0$).
Secondly, the leading order correction to the threshold from the rotation is quadratic in the spin parameter
with a positive ${\cal O}(1)$ coefficient.
Again, this is consistent with the naive expectation that more gravitational force, namely larger amplitude of the density
contrast, is needed for the overdense region to undergo the gravitational collapse against rotation
which effectively produces repulsive force.
Although this result is obtained under the various approximations and assumptions,
we expect from the physical ground that the quadratic dependence of the threshold 
(\ref{threshold-spin}) on $a_K$ with ${\cal O}(1)$ positive coefficient is a generic qualitative feature and
more precise computation will change the magnitude of the coefficient.
Indeed, the analysis in the following section which is based on the recent refined argument \cite{Harada:2013epa} shows that
this is the case.

\section{New estimation of the threshold of the PBH formation}
The argument developed in the previous section is simple and intuitive for deriving the formation threshold of PBHs.
The basic idea was the comparison between the size of the overdense region and the Newtonian Jeans length
at the turnaround time, which was initially used in \cite{Carr:1975qj} to derive the threshold for non-rotating PBHs.
Recently, in \cite{Harada:2013epa}, a new threshold was derived based on a refined criterion from the previous one.
In short, the new criterion is a comparison between the size of the overdense region and the 
traveling distance of the sound wave by the time of the maximum expansion on the cosmological background,
which clarifies the $O(1)$ ambiguity inherent to the Jeans length used in the previous section by going beyond
the simple Newtonian Jeans length.
It was demonstrated that the new criterion yields a threshold which shows better agreement with the ones
obtained by numerical simulations \cite{Harada:2013epa}.
In light of this situation, it is natural to adopt the refined criterion to derive the threshold of 
the rotating PBHs.
Motivated by this consideration, 
we derive the new threshold based on the refined criterion in this section.

As in the previous section, we consider that the overdense region which later forms a PBH is described 
by a part of the closed Friedmann universe whose metric is given by 
\begin{equation}
    ds^2=g_{\mu \nu}^{(0)}dx^\mu dx^\nu=-dt^2+a^2(t) (d\chi^2+\sin^2 \chi d\Omega^2 )=-dt^2+a^2(t) \gamma_{ij} dx^idx^j, \label{metric-closedFLRW}
\end{equation}
where $\gamma_{ij}$ is the 3-dim metric given by
\begin{equation}
    \gamma_{ij}=\left(
    \begin{array}{ccc}
	    1 & 0 & 0 \\
    	0 & \sin^2 \chi & 0 \\
    	0 & 0 & \sin^2 \chi \sin^2 \theta
    \end{array}
    \right)~.
\end{equation}
The overdense region is covered by $0 \le \chi \le \chi_a$, where
$\chi_a$ is eventually related to the amplitude of the density contrast at the horizon reentry \cite{Harada:2013epa}. 
The evolution equations of this universe are given by 
\begin{equation}\label{bg-1}
    \frac{1+\da^2}{a^2}=\frac{8\pi G}{3}\rho~,~~~~~1+\da^2+2a{\ddot a}=-8\pi G a^2 P~,
\end{equation}
from which we obtain
\begin{equation}
    (1+3w) (1+\da^2)+2a{\ddot a}=0~,
\end{equation}
where $ \rho $ and $ P $ are, respectively, the energy density and the hydrostatic pressure of the perfect fluid within the overdense region, 
and the equation of state parameter is then $ w=P/\rho =c_s^2$. 
After the overdense region re-enters the Hubble horizon, it continues to expand until reaching the maximum expansion,
and then it shrinks and finally collapses into a BH. At the time of the maximum expansion, we have 
\begin{equation}
	\frac{1}{a^2_{\rm max}} =\frac{8\pi G}{3} \rho_{\rm max}
\end{equation}
where the subscript denotes the quantities evaluated at the maximum expansion. 

\subsection{Modeling of the rotation}
\label{model-rot}
Now we consider adding small rotation on top of the overdense region.
Mathematically, this amounts to defining the rotational perturbation on the closed FLRW metric.
In the context of the cosmological perturbation theory,
this perturbation belongs to the so-called vector-type perturbation.
Without its source, this type of perturbations is known to decay as the Universe expands.
The realistic source of giving the rotation is, for instance, the tidal force generated by the surrounding inhomogeneities \cite{DeLuca:2019buf}.
By this source, the overdense region initially having vanishing rotation on super-Hubble scales acquires angular momentum upon the horizon reentry. 
In what follows, we consider the vector perturbation which remains to be vanishing while the overdense region is on super-Hubble
and is instantly generated at the horizon reentry.
Generally, the generated velocity distribution would have a complicated configuration, whose modeling is beyond the scope of this paper. 
Instead, we restrict ourselves to the axisymmetric perturbation in this paper.
There are two reasons for doing this.
Firstly, it is physically the simplest type of perturbations that represents the rotation of the system.
We expect that the essential physical effects of rotation would be already observed in this simple case.
Secondly, because of its simplicity of the perturbation,
evaluation of its effect on the background expansion can be done analytically
and this helps us to understand the results intuitively. 
Based on these considerations, it is a natural first step to adopt the axisymmetric perturbations
to understand physically the effects of rotation on the PBH formation.

Even after we restrict our analysis to the axisymmetric perturbation, 
there are still infinitely many degrees of freedom to describe the rotation as we can see when doing multipole expansion. 
Therefore, for the picture to be simple and clear, we consider dipolar and axisymmetric rotation.
In terms of the perturbation theory on spherically symmetric background, 
such perturbation is described by the odd parity mode with the spherical harmonic with $(\ell,m)=(1,0)$ \cite{Regge:1957td}. 
If we write down the perturbations of metric and the fluid on the closed FLRW metric as
\begin{align}
	ds^2 &=\left( g^{(0)}_{\mu\nu} +h_{\mu\nu} \right) dx^{\mu}dx^{\nu} ~, \\
	u_{\mu}&=\bar{u}_{\mu}+\delta u_{\mu} ~,
\end{align}
the non-vanishing components in the present case are then given by 
\begin{equation}
    h_{t\phi}=h_0 (t,\chi) \sin^2 \theta~,~~~\delta u_\phi=V(t,\chi) \sin^2 \theta ~. \label{pert-rotation}
\end{equation}
Here we have chosen a gauge in which $h_{r \phi}=0$.
For these perturbations, it is straightforward to obtain 
\begin{equation}
    \delta u^\phi=\frac{1}{a^2 \sin^2 \chi} (V-h_0)~. \label{eq-for-uphi}
\end{equation}
In terms of cosmological perturbation theory, the vector-type metric perturbations are defined as 
\begin{equation}
    h_{tt}=0~,~~~h_{t i}=B_i~,~~~h_{ij}=\nabla_i E_j+\nabla_j E_i~,~~~\delta u_t=0~,~~~\delta u_i=V_i~, \label{def-vector}
\end{equation}
where $\nabla_i$ is the covariant derivative with respect to the spatial metric $\gamma_{ij}$,
and the 3-vectors $B_i,~E_i,~V_i$ satisfy the transverse condition
\begin{equation}
    \nabla^i B_i=\nabla^i E_i=\nabla^i V_i=0~.
\end{equation}
By comparison between Eq.~\eqref{pert-rotation} and \eqref{def-vector}, we can relate the perturbation variables to $B_i,~E_i,~V_i$ as 
\begin{equation}
	B_{\phi} = h_0(t,\chi)\sin^2\theta ~,~~~ V_{\phi} =V(t,\chi)\sin^2\theta ~.
\end{equation}
where only the non-vanishing components are presented. 
It is straightforward to confirm that they do satisfy the transverse conditions. 
Having defined the rotational perturbations, the next step is to derive the equations of motion of the perturbations and solve them. 

\subsection{Shape of the rotational perturbations}

For later convenience, we introduce a function $ h(t,\chi) $ as $ h=ah_0 $.
Then, the linearized Einstein equations end up with the following set of two equations 
\begin{align}
    2\frac{\cos \chi}{\sin \chi} {\dot h}-{\dot h}' &=0~, \label{odd-Eineq-1} \\
    2 \left( 1-\frac{\cos^2 \chi}{\sin^2 \chi} \right) h+h''&=6 (1+w) (1+{\dot a}^2) aV ~. \label{odd-Eineq-2}
\end{align}
The general solution of Eq.~\eqref{odd-Eineq-1} is given by
\begin{equation}
    h(t,\chi)=C(t) \sin^2 \chi+J(\chi)~, \\
\end{equation}
where $ C(t) $ and $ J(\chi) $ are arbitrary functions of $ t $ and $ \chi $, respectively. 
There are residual gauge degrees of freedom in this solution. 
Actually, under the transformation $ x^\mu \to {\tilde x}^\mu=x^\mu+\xi^\mu $ given by \footnote{Here $ a,b=2,3 $ represents 
the angular directions $ \theta $ and $ \phi $. 
The definition of $ E_{ab} $ is $ E_{ab}\equiv \sqrt{\det\gamma} \epsilon_{ab} $ where $ \epsilon_{\theta\phi}=1 $ and $ \gamma_{ab}={\rm diag}(1,\sin^2\theta) $.}
\begin{equation}
    \xi_a=E_a^{~b} \partial_b \Lambda~,~~~~~\Lambda=C_1(t) \sin^2 \chi Y_{10}~,
\end{equation}
where $ C_1(t) $ is an arbitrary function of $ t $, all the components except for $ h_{ta} $ and $ u_a $ remain zero, and $ h_0 $ and $ V $ transform as
\begin{equation}
    h_0 \to h_0+(-{\dot C_1}+2H) \sin^2 \chi~,~~~~~V \to V~.
\end{equation}
Thus, the first term of $ h(t,\chi) $ is a gauge mode, and we can set $ C=0 $ without a loss of generality.
On the other hand, $ J(\chi) $ cannot be set to zero arbitrarily since it is determined by the radial distribution of the physical angular momentum. 
Substituting $ h(t,\chi)=J(\chi) $ into Eq.~\eqref{odd-Eineq-2}, we obtain an equation for $ V $ and $ J $ as
\begin{equation}
    2 \left( 1-\frac{\cos^2 \chi}{\sin^2 \chi} \right) J(\chi)+J''(\chi)=6 (1+w) (1+{\dot a}^2) aV(t,\chi) ~. \label{eq-J-1}
\end{equation}
In order for this equation to be consistent, the right hand side should not depend on $ t $. Using the Friedmann equation, this consistency condition fixes $ V(t,\chi) $ as
\begin{equation}
    V(t,\chi)=\left( \frac{a}{a_{\rm max}} \right)^{3w} V_{\rm max}(\chi)~,
\end{equation}
where $V_{\rm max}(\chi)$, which represents the gauge invariant radial distribution of angular velocity of the fluid, is fixed by the initial condition.  
For instance, $V_{\rm max} (\chi) \propto \sin^2 \chi$ corresponds to a uniform rotation in the absence of metric perturbation (see Eq.~\eqref{eq-for-uphi}).
Plugging this back into Eq.~\eqref{eq-J-1}, we obtain
\begin{equation}
    2 \left( 1-\frac{\cos^2 \chi}{\sin^2 \chi} \right) J(\chi)-6 (1+w) a_{\rm max}V_{\rm max} (\chi)+J''(\chi)=0~. \label{eq-J-2}
\end{equation}
Requiring the regularity condition at $\chi=0$ as $J(0)=J'(0)=0$, the solution of this equation is given by
\begin{equation}
    J(\chi)=6(1+w) a_{\rm max}\sin^2 \chi \int_0^\chi \frac{d\chi'}{\sin^4 \chi'} \int_0^{\chi'} d\chi''~\sin^2 \chi'' V_{\rm max}(\chi'')~. \label{sol-J}
\end{equation}
From Eq.~\eqref{sol-J}, we find that $J$ generically diverges at $\chi=\pi$ unless the form of $V_{\max}(\chi)$ is fine-tuned adequately.
In particular, if the sign of $J_{\rm max}$ is the same everywhere, the divergence is inevitable.
This means that we cannot add the non-zero angular momentum (at least at the linear level) to the closed FLRW Universe \cite{Lyndel:1995jm}.
This singularity is not problematic in the present analysis since the overdense region which we approximate as the closed FLRW universe
covers only a fraction of the entire closed universe which is parametrized by $\chi_a$.

To summarize, the rotational perturbations that we consider can be written as
\begin{equation}
h_{t\phi}=\frac{a_{\rm max}}{a(t)} j(\chi) \sin^2 \theta,~~~~~~~
\delta u_\phi=\left( \frac{a(t)}{a_{\rm max}} \right)^{3w} V_{\rm max}(\chi) \sin^2 \theta, \label{v-perturbation}
\end{equation}
where $j(\chi) \equiv J(\chi)/a_{\rm max}$.
	
\subsection{Effective energy-momentum tensor and backreaction}
As is mentioned at the beginning of this section,
the basic idea of \cite{Harada:2013epa} is the comparison between the size of the overdense
region and the sound traveling distance by the time of the maximum expansion.
When rotation is added, it is expected that rotation energy of the fluid increases the total energy density appearing in the Friedmann equation
and changes the evolution of the scale factor of the overdense region from the non-rotating case.
This will delay the maximum expansion time and allow the sound wave to travel over a longer distance.
As a result, the sound wave will prevent the overdense region from collapsing even for the density contrast 
which will lead to PBH formation without rotation.
The purpose of this and next subsections is to embody this physical picture in the mathematical language.

The effects of rotation on the expansion of the overdense region are described as backreaction which appears at second order in the rotational perturbations.
In \cite{Abramo:1997hu}, a method was proposed to construct gauge-invariant effective energy-momentum tensor at second order
in perturbations.
The key point is to introduce quantities by which the gauge-invariance at the backreaction level is ensured.
In our present case where only the vector-type perturbations exist at the linear order,
the method in \cite{Abramo:1997hu} shows that one can directly use the perturbations given by Eq.~(\ref{v-perturbation}) for
computing the second order Einstein equations. 
Formally, the Einstein equation for the background with second order perturbations included can be written as \footnote{
We ignore the effects of scalar perturbations sourced by the rotational perturbation by the second-order effect.}
\begin{align}
    G^{\rm (0)}_{\mu\nu} =8\pi G T^{\rm (0)}_{\mu\nu}+8\pi G \left(-\frac{1}{8\pi G} \left<G^{\rm (2)}_{\mu\nu}\right> +\left< T^{\rm (2)}_{\mu\nu}\right> \right)  ~.
\end{align}
The part inside the parenthesis on the right-hand side can be regarded as an effective energy-momentum tensor sourced by the angular momentum. 
Its time-time component is the contribution to the energy density
\begin{align}\label{deltarho-original}
    \Delta\rho (t)= -\frac{1}{8\pi G} \left<G^{\rm (2)}_{00}\right> +\left< T^{\rm (2)}_{00}\right> ~.
\end{align}
which will affect the evolution of the closed FLRW universe. 
According to the previous results in Eq.~\eqref{v-perturbation}, the time-time component of the energy-momentum tensor is given by 
\begin{align}
    T^{\rm (2)}_{00}= \frac{1+w}{a^4} \frac{\sin^2\theta}{\sin^2\chi} \bar{\rho} \left( a_{\rm max} j -\frac{a^{1+3w}}{a^{3w}_{\rm max}} V_{\rm max} \right)^2
\end{align}
where we have taken into account the normalization condition $ u^{\mu}u_{\mu}=-1 $ to the second order. 
Similarly, the Einstein tensor is given by 
\begin{align}
    G^{\rm (2)}_{00}&= \frac{a^2_{\rm max}\sin^2\theta}{4a^6\sin^2\chi} \left[ 4\left( -3+3(1+w) +\frac{1}{\sin^2\chi} + 3w \da^2 \right)j^2 -j'^2 +4\left( j'\cot\chi- j'' \right)j \right]~.
\end{align}
By taking the spatial average of a quantity over the closed FLRW background as 
\begin{align}
    \left<{\cal O}\right> =\frac{ \int_V {\cal O} \sin^2{\chi} \sin{\theta} d\chi d\theta d\varphi }{\int_V \sin^2{\chi} \sin{\theta} d\chi d\theta d\varphi} ~,
\end{align}
we obtain the spatially averaged second-order energy-momentum tensor and Einstein tensor as follows 
\begin{align}
    \left<T^{\rm (2)}_{00}\right> =& \frac{4 (1+w)\bar{\rho}}{3 a^4(\chi_a-\sin\chi_a\cos\chi_a)}\int^{\chi_a}_{0} d\chi\left( a_{\rm max} j -\frac{a^{1+3w}}{a^{3w}_{\rm max}} V_{\rm max}\right)^2~, \label{too} \\
    \left<G^{\rm (2)}_{00}\right> =& -\frac{ a^2_{\rm max}}{3a^6 (\chi_a-\sin\chi_a\cos\chi_a)} \times \nonumber \\
    &\int^{\chi_a}_{0} d\chi \left[ (j'-2j\cot\chi)^2  +24(1+w) j V_{\rm max} -12( 1+w+w \da^2)j^2 \right] \label{goo}
\end{align}
where we have used Eq.~\eqref{sol-J}. 

So far, we have not specified the radial configuration of the angular velocity,
and Eqs.~(\ref{too}) and (\ref{goo}) are valid for any function $j(\chi)$.
In what follows, in order to obtain the concrete expression of the threshold, 
we consider $ V_{\rm max} (\chi) =V_f \sin^2\chi $ where $ V_f={\rm const} $, 
which represents a uniform rotation of the fluid in the absence of the metric perturbation. 
For this radial distribution, \eqref{sol-J} yields 
\begin{align}
    j(\chi)= \frac{3}{4}(1+w)V_f \left[ \frac{5}{3}\sin^2\chi +1-\chi\left(\sin (2\chi) +\cot\chi \right) \right]~.
\end{align}
Direct substitution of this function to Eqs.~(\ref{too}) and (\ref{goo}) gives complicated functions of $\chi_a$.
Instead, we treat $\chi_a$ as a small parameter and compute $\left<T^{\rm (2)}_{00}\right>$ and $\left<G^{\rm (2)}_{00}\right>$ 
perturbatively although $\chi_a$ for the case of the radiation fluid ($\chi_a=\sqrt{3}\pi/6 \approx 0.91$) is not so small.
To the fourth order in $\chi_a$, we obtain
\begin{align}
    \left<T^{\rm (2)}_{00}\right> &= \frac{2(1+w)V^2_f \bar{\rho}a^{6w-2}}{5a^{6w}_{\rm max}}  \chi^2_a -\frac{2(1+w)V^2_f \bar{\rho}a^{3(w-1)}}{525a^{3w}_{\rm max}}\left[ 90(1+w) a_{\rm max} +29\frac{a^{3w+1}}{a^{3w}_{\rm max}} \right]\chi^4_a ~, \\
    \left<G^{\rm (2)}_{00}\right> &= -\frac{198(1+w)^2 a^2_{\rm max}V^2_f}{175a^6}\chi^4_a ~.
\end{align}
Plugging them back to Eq.~\eqref{deltarho-original}, we obtain the effective energy density 
$ \rho_{\rm eff}=\bar{\rho}+\Delta \rho $, 
where $ \Delta\rho $ is the backreaction from the second-order vector-type perturbations given by 
\begin{align}\label{deltarho-1}
    \Delta \rho =& \frac{2(1+w)V^2_f \bar{\rho} a^{6w-2}}{5 a^{6w}_{\rm max}} \chi^2_a  \nonumber \\
    &+ \frac{(1+w)^2V^2_f}{2100\pi G a^6} \left[ 297a^2_{\rm max} -8\pi G \frac{a^{3(w+1)}}{a^{3w-1}_{\rm max}} \left( 90+\frac{29}{(1+w)}\frac{a^{3w+1}}{a^{3w+1}_{\rm max}} \right)\bar{\rho} \right] \chi^4_a  ~.
\end{align}
From this equation, we find that relative magnitude of the ${\cal O}(\chi_a^4)$ term 
to the ${\cal O}(\chi_a^2)$ for the case of the radiation fluid is given 
\begin{equation}
\frac{\Delta \rho^{(4)}}{\Delta \rho^{(2)}}=\frac{1}{105} \left( -29+12 \frac{a_{\rm max}^2}{a^2} \right) \chi_a^2.
\end{equation}
For $a_{\rm hc} \le a \le a_{\rm max}$, magnitude of this ratio is less than 20$\%$.
Thus, in the following analysis, we ignore $\Delta \rho^{(4)}$ and keep only the ${\cal O}(\chi_a^2)$ term
in $\Delta \rho$.

\subsection{Threshold of the PBH Formation}

For notational simplicity, we write $\Delta \rho$ given by Eq.~(\ref{deltarho-1}) up to ${\cal O}(\chi_a^2)$ as
\begin{equation}
	\Delta\rho =\epsilon_0 \bar{\rho} a^{6w-2}
\end{equation}
where $\epsilon_0$ is defined by
\begin{equation}
    \epsilon_0 \equiv \frac{2(1+w)}{5a^{6w}_{\rm max}} V^2_f \chi^2_a ~.
\end{equation}
In what follows, we treat $\epsilon_0$, which is second order in the rotation, as a small quantity.
As is explained in \ref{model-rot}, we assume that the rotational perturbations are generated upon the horizon reentry.
Thus, $\Delta \rho$ is present only after the horizon reentry, i.e. $a > a_{\rm hc}$.
In order to reflect this feature, we promote $ \epsilon_0 $ to a time-dependent quantity $\epsilon$ as 
\begin{equation}
    \epsilon (a) =\epsilon_0 \Theta(a-a_{hc}) 
\end{equation}
where $ \Theta(a) $ is the step function. Thus, $\Delta \rho$ we use is actually given by $\epsilon \bar{\rho} a^{6w-2}$. 
In the uniform Hubble gauge, $ a_{hc} $ is defined as \cite{Harada:2013epa}
\begin{equation}
    a_{hc} \sin \chi_a =H_b^{-1}(t_{hc})=H^{-1}(t_{hc})=\left. \frac{a}{\dot{a}} \right \vert_{t=t_{hc}}
\end{equation}
where $ H_b $ is the flat FLRW background around the overdense region. 
With $\Delta \rho$, the Friedmann equation including the rotational effects is given by
\begin{equation}\label{correctedfried}
  \frac{1+\dot{\tilde{a}}^2}{\tilde{a}^2} = \frac{8\pi G}{3} \bar{\rho}(\tilde{a}) (1+\epsilon (\tilde{a}) \tilde{a}^{6w-2} ), 
\end{equation}
where we denote the scale factor as $ \tilde{a} $ to emphasize that it evolves under the rotational effects.
At the maximal expansion time, we have 
\begin{equation}\label{newmax:1}
    \frac{1}{\tilde{a}^2_{\rm max}} = \frac{8\pi G}{3} \bar{\rho}(\tilde{a}_{\rm max}) (1+\epsilon \tilde{a}^{6w-2}_{\rm max}). 
\end{equation}
Notice that $ \bar{\rho}(\tilde{a}_{\rm max}) $ is generally different from $ \bar{\rho}_{\rm max} =\bar{\rho}(a_{\rm max})$ 
since $a$ and ${\tilde a}$ evolve differently after the horizon reentry. Indeed, we have
\begin{align}\label{newrhomax-1}
    \bar{\rho} (\tilde{a}) =\bar{\rho}_{\rm max} \left( \frac{a_{\rm max}}{\tilde{a}_{\rm max}} \right)^{3(1+w)} \left( \frac{\tilde{a}_{\rm max}}{\tilde{a}} \right)^{3(1+w)} \equiv \bar{\rho}(\tilde{a}_{\rm max}) \left( \frac{\tilde{a}_{\rm max}}{\tilde{a}} \right)^{3(1+w)} ~.
\end{align}

In order to derive the threshold of the PBH formation along the same approach as \cite{Harada:2013epa},  
we need to compare the sound crossing distance with the maximum size of the overdense region.
The sound crossing time is given by
\begin{equation}
	\sqrt{1/w} \int^{\chi_s}_0 d\chi = \int^{t_{\rm max}}_0 \frac{dt}{\tilde{a}}= 
	\int^1_0 \frac{du}{u\sqrt{\frac{1+\epsilon(\tilde{a}_{\rm max}u)\tilde{a}^{6w-2}_{\rm max} u^{6w-2}}{1+\epsilon \tilde{a}^{6w-2}_{\rm max}} \frac{1}{u^{1+3w}}-1}},
\end{equation}
where in the last equation we changed the variable $ u\equiv \tilde{a}/\tilde{a}_{\rm max} $. 
Expanding the integrand up to first order in $\epsilon$, we can perform the integration over $u$.
The result is given by 
\begin{equation}
    \sqrt{1/w} \chi_s = \frac{\pi}{1+3w} +\frac{\epsilon_0 a^{6w-2}_{\rm max}}{2} \frac{2\left( a_{hc}/a_{\rm max}\right)^{3w/2}}{(1+3w)\sqrt{ a_{\rm max}/a_{hc}-\left( a_{hc}/a_{\rm max} \right)^{3w}}} +\frac{\epsilon_0 a^{6w-2}_{\rm max}}{2} \Xi \left(\frac{a_{hc}}{a_{\rm max}},w \right) \label{chis}
\end{equation}
where $ \Xi (x,\alpha) $ is given by 
\begin{align}
    \Xi (x,\alpha) \equiv& \frac{1}{3(\alpha -1)} \Bigg[ \frac{2\sqrt{\pi}\Gamma\left( \frac{15\alpha-3}{6\alpha+2} \right)}{\Gamma\left( \frac{3\alpha-3}{3\alpha+1} \right)} +\frac{\sqrt{x^{-1-3\alpha}-1}}{(1+3\alpha)(x^4-x^{3-3\alpha})} \Bigg( 5-9\alpha +6(\alpha-1)x^4 \nonumber \\ 
    & +(1+3\alpha)x^{1+3\alpha} +(9\alpha-5)\sqrt{1-x^{1+3\alpha}} _2F_1\left( \frac{1}{2},\frac{3\alpha-7}{6\alpha+2};\frac{9\alpha-5}{6\alpha+2}; x^{1+3\alpha} \right) \Bigg) \Bigg] ~.
\end{align}
Above, $ \Gamma(z) $ is the gamma function and $ _2F_1 (a,b;c;z) $ is the hypergeometric function. 
Without rotation, $a_{\rm max}$ is related to $a_{\rm hc}$ as \cite{Harada:2013epa}
\begin{equation}\label{ahc:w}
    \frac{a_{hc}}{a_{\rm max}} =\frac{1}{(2+\cot^2\chi_a)^{\frac{1}{1+3w}}} ~.
\end{equation}
Using this relation, Eq.~(\ref{chis}) becomes
\begin{equation}\label{chi_s}
	\chi_s = \frac{\pi\sqrt{w}}{1+3w} +\frac{\sqrt{w}\epsilon_0 a^{6w-2}_{\rm max}}{1+3w} \sin\chi_a + \frac{\sqrt{w}\epsilon_0 a^{6w-2}_{\rm max}}{2} \Xi \left( \frac{1}{(2+\cot^2\chi_a)^{\frac{1}{1+3w}}},w \right) ~.
\end{equation}
Having obtained the sound crossing distance, we can now derive the threshold of PBH formation 
by requiring $ \chi_a >\chi_s $ which gives $ 1\geq \delta^{\rm UH}_{H}\equiv\sin^2\chi_a >\delta^{\rm UH}_{Hc} \equiv \sin^2\chi_s $ \cite{Harada:2013epa}\footnote{Here we only consider the type-I fluctuations with $ 0<\chi_a<\pi/2 $. The detailed discussion about the type-I and -II fluctuations are presented in \cite{Kopp:2010sh}.} so that 
\begin{align}
	\delta^{\rm UH}_{Hc} \simeq& \sin^2\left( \frac{\pi\sqrt{w}}{1+3w} \right) \nonumber \\
	&+\epsilon_0 \left[ \frac{\sqrt{w} a^{6w-2}_{\rm max}}{1+3w} \sin\chi_a + \frac{\sqrt{w}a^{6w-2}_{\rm max}}{2} \Xi \left( \frac{1}{(2+\cot^2\chi_a)^{\frac{1}{1+3w}}},w \right) \right] \sin\left( \frac{2\pi\sqrt{w}}{1+3w} \right)
\end{align}
up to $ {\cal O}(\epsilon_0) $. 
This should be written in terms of the angular momentum or, equivalently, the dimensionless spin parameter. 
As is the case for Sec.~\ref{simple}, we assume that the angular momentum is conserved throughout the gravitational collapse after the maximum expansion
and the initial angular momentum of the resultant BH is equal to the angular momentum of the overdense region. 
According to the result in Appendix~\ref{appendix-killing}, we have 
\begin{equation}
	J=\frac{8(1+w)\pi a^{3(w+1)}\bar{\rho}}{15a^{3w}_{\rm max}}V_f \chi^5_a
\end{equation}
up to leading order in $ \chi_a $ expansion. 
By substituting $J $ for $ V_f $ in $ \epsilon_0 $, we have 
\begin{equation}
	\epsilon_0 =\frac{5}{2(1+w)a^{6w}_{\rm max}\chi^2_a} \frac{J^2}{M^2_{\rm BH}},
\end{equation}
where we have defined the mass of the PBH to be $ M_{\rm BH}\equiv 4\pi a^3_{\rm max}\bar{\rho}_{\rm max} \chi^3_a/3 $.
Thus, the threshold of the density contrast for the rotating PBH (on the uniform Hubble slice) becomes 
\begin{equation}
	\delta^{\rm UH}_{Hc} =  \sin^2\left( \frac{\pi\sqrt{w}}{1+3w} \right) + \frac{5\sqrt{w}}{2(1+3w)(1+w)a^2_{\rm max}\chi^2_a} \frac{J^2}{M^2_{\rm BH}}\sin\left( \frac{2\pi\sqrt{w}}{1+3w} \right) \left[ \sin\chi_a +\frac{1+3w}{2} \Xi \right] \label{delta-generalw}~.
\end{equation}

Now let us investigate the most interesting case where the fluid is radiation. 
For radiation, $ \Xi (x,1/3)=0 $, and the above expression is simplified as
\begin{equation}\label{delta-radw}
	\delta^{\rm UH}_{Hc} = \sin^2\left( \frac{\sqrt{3}\pi}{6} \right)+ 
	\frac{5\sqrt{3}}{64} \sin\left( \frac{\sqrt{3}\pi}{3} \right) 
	\left( \frac{2GM_{\rm hc}}{a_{\rm hc} \chi_a} \right)^2 
	\frac{\sin\chi_a}{(2+\cot^2\chi_a)^2} a_K^2,  
\end{equation}
where we have replaced $J$ by the spin parameter $a_K= J/(G M^2_{\rm BH}) $.

In the slow-rotation limit, we can replace $\chi_a$ appearing in the coefficient in front of $a_K^2$
by the value in the non-rotating case, i.e. $\sqrt{3}\pi/6$. 
Assuming $2GM_{\rm hc}/(a_{\rm hc}\chi_a)=1$, we finally obtain
\be
    \delta^{\rm UH}_{Hc} \simeq 0.62+ 
    0.015a^2_K ~. \label{fdelta}
\ee
This is the main result of this work. In order to link with the primordial density fluctuations in the inflationary power spectrum, we here transform this threshold in the uniform Hubble gauge into the one in comoving density gauge. Given Eq.~\eqref{chi_s}, the calculation is straightforward \cite{Kopp:2010sh}, the averaged fluctuation 
\begin{align}
	\bar{\zeta}_c&=\frac{1}{3} \ln \left[ \frac{3\chi_s-3\sin\chi_s\cos\chi_s}{2\sin^3\chi_s} \right] \\
	&\simeq 0.09+0.003 a^2_K 
\end{align}
and the curvature perturbation 
\begin{align}
	\zeta_c&\simeq -2 \ln \cos \frac{\chi_s}{2} \\
	&\simeq 0.21+0.008 a^2_K 
\end{align}
which are given at the horizon crossing. 

\section{Discussions}
In the previous section, by adopting the physical criterion of the PBH formation used in \cite{Harada:2013epa},
we have derived the threshold of the density contrast for the formation of a rotating PBH (i.e. Eq.~(\ref{delta-radw})
or Eq.~(\ref{fdelta})).
As is explained in the previous section, there are several crucial assumptions and approximations used to derive this result.
They include modeling of the rotation as axially symmetric and uniform vectorial perturbations on the closed FLRW metric, 
sudden creation of such rotation by the surrounding matter inhomogeneities upon horizon reentry,
evaluation of effects of rotation as a backreaction to the closed FLRW metric, 
approximation by neglecting contributions from higher order terms in $ \chi_a $,
and ignoring the effects of scalar perturbations sourced by the rotational perturbation by the second-order effect.
Evaluating how much the inclusion of those effects changes our result is a non-trivial task and beyond the scope of this paper.
Given that there is already about a factor of 5 difference (in terms of $a_{K, \rm max}$) between the result based on the 
Newtonian picture in Sec.~\ref{simple} and the one in the last section,
the same level of changes may be made after those effects are included in the analysis.
The most straightforward way to clarify this issue would be to do numerical simulations with initial conditions
determined by the statistical properties of initial primordial perturbations.
All the existing simulations assume spherical symmetry \cite{Niemeyer:1999ak, Shibata:1999zs, Polnarev:2006aa, Musco:2008hv, 
Musco:2012au,Nakama:2013ica, Escriva:2019nsa} 
and it is a very interesting challenge to perform
simulations that include rotation.

Both our analyses based on the original picture \cite{Carr:1975qj} and on the recent refined version \cite{Harada:2013epa} 
show that the contribution to the threshold $\delta_{\rm th}(J)$ from $J$ is quadratic in $J$ with positive coefficient.
This result is natural given the physical ground that rotation impedes gravitational collapse.
We expect that this feature remains true even after more precise calculations that include the ignored 
effects mentioned above are done.

As we discussed in the Introduction, there are a couple of factors that determine the spin distribution of PBHs.
Assuming that the probability distribution $P(\delta_M, J)$ defined in Eq.~(\ref{intro1}) is a monotonically decreasing 
function of $\delta_M$ and $J$ such as the case for the Gaussian primordial perturbations, the typical value of the initial spin of PBHs 
will be primarily determined by 
the competition between two aspects whichever is smaller, (1) the standard deviation of the angular momentum distribution of the 
primordial density perturbations, and (2) the maximum spin parameter obtained by combining $\delta_{\rm th}$
with the probability density $P(\delta_M, J)$.
It was argued that the typical value of the angular momentum of the Gaussian perturbations is $a_K ={\cal O}(0.01)$ \cite{Mirbabayi:2019uph, DeLuca:2019buf}.
This is much smaller than the maximum spin parameter derived in this paper.
This may suggest that, at least in the case of Gaussian perturbations, 
the initial spin of PBHs is essentially determined by the angular momentum distribution of the primordial perturbations, 
and that the difference of $\delta_{\rm th}$ for different angular momentum does not play a significant role in shaping the spin distribution of PBHs.
Yet, this picture may not hold for strongly non-Gaussian perturbations.

\section{Conclusion and Outlook}

In this paper, we considered the formation of PBHs by direct gravitational collapse of large primordial density perturbations 
with angular momentum in the radiation-dominant epoch. 
The formation rate, in this case, is suppressed by the rotational effect compared with the non-rotating case, 
which can be seen from the enhancement of the threshold $ \delta_{\rm th} $ as the angular momentum increases. 
The main result is shown in Eq.~\eqref{delta-radw}. 

We modeled the rotation of the overdense region by vector-type cosmological perturbations which could be, for example, 
generated by the inhomogeneity of the density profile around the peak \cite{DeLuca:2019buf}. 
Without a loss of the essence of physics, we restricted our analysis to the dipolar and axisymmetric rotation for technical simplicity. 
We further assumed that the angular momentum of the overdense region is only switched on soon after the horizon reentry 
and it is conserved after the beginning of the collapse. 
Under these assumptions, we included the rotational effect as the correction to the energy-momentum tensor, 
backreacting on the evolution of the closed FLRW background. 
Specifically, the angular momentum modifies the energy density, 
which results in modification of the sound traveling distance within the overdense region 
such that the formation threshold of BHs, $ \delta_{\rm th} $, is enhanced by an additional contribution quadratic in $ a_K $ up to leading order. 
In other words, the rotation suppresses the formation of PBHs as second order in the spin parameter. 
Thus, the larger the spin is, the less productive to form a BH. 
The angular momentum of the BHs at the formation time is therefore rather small. 

Some of the BH merger events detected by LIGO/Virgo have small or vanishing magnitudes of
the effective aligned spin \cite{LIGOScientific:2018mvr}.
As more gravitational wave events will be detected in the future, 
more precise determination of the spin distribution will become possible so that the results in this paper can be contrasted with the observations.

\section*{Acknowledgement}

The authors thank B. Carr, L. Hui, M. Takada, and J. Yokoyama for useful discussion. 
MH was supported by the Global Science Graduate Course (GSGC) program of the University of Tokyo and the JSPS Research Fellowships for Young Scientists. 
TS is supported by JSPS Grant-in-Aid for Young Scientists (B) No.15K17632, by the MEXT Grant-in-Aid for Scientific Research on In-novative Areas 
No.15H05888, No.17H06359, No.18H04338, and No.19K03864.

\appendix 

\section{Killing Vector}\label{appendix-killing}

In order to find out a relation between the angular momentum of the overdense region and the threshold $ \delta^{\rm UH}_{Hc} $, 
we first figure out the relation between angular momentum and rotational velocity. 
According to our set up, the overdense region has axial symmetry, which means that we can find a Killing vector $ \xi^{\mu}=\partial_{\phi} $ that satisfies 
\begin{equation}
    {\mathcal L}_{\xi} g_{\mu\nu}=0 
\end{equation}
from which we can obtain the lower index Killing vector 
\begin{equation}
     \xi_{\mu} =g_{\mu\nu} \xi^{\nu} =g_{\mu\phi} =\left( \frac{a_{\rm max}}{a}j(\chi)\sin^2\theta~,~0~,~0~,~a^2\sin^2\chi \sin^2\theta \right)
\end{equation}
Define $ j^{\mu}\equiv T^{\mu\nu}\xi_{\nu} $ where $ T^{\mu\nu} $ is the energy-momentum tensor which is conserved. 
It is easy to prove that the divergence of $j^{\mu} $ vanishes, 
\begin{align}
    \nabla_{\mu} j^{\mu} =\nabla_{\mu} \left( T^{\mu\nu}\xi_{\nu} \right) &=\xi_{\nu} \nabla_{\mu} T^{\mu\nu} +T^{\mu\nu} \nabla_{\mu} \xi_{\nu} \\
    &=0+ \frac{1}{2} T^{\mu\nu} \left( \nabla_{\mu} \xi_{\nu} +\nabla_{\nu} \xi_{\mu} \right) =0
\end{align}
so that we have
\begin{align}
\partial_0 \left( \sqrt{-g} j^0 \right) &=-\partial_i \left( \sqrt{-g} j^i \right) ~.
\end{align}
Therefore, we can define a charge $J(t) $ as 
\begin{equation}
  J(t) \equiv \int_{\Sigma_t} d^3x \sqrt{-g} j^0
\end{equation}
where $ \Sigma_t $ is a $ t=\text{const} $ space-like slice. 
We will take it as the closed Friedmann universe. 
The time derivative of $J$ is 
\begin{align}
    \dot{J}(t) &=\int_{\Sigma_t} d^3x \partial_0 \left( \sqrt{-g} j^0 \right) \\
    &= -\int_{\Sigma_t} d^3x \partial_i \left( \sqrt{-g} j^i \right) \\
    &= -\oint_{\partial \Sigma_t} \sqrt{-g} \vec{j}\cdot d\vec{S} ~.
\end{align}
This vanishes since the flux ${\vec j}$ is not passing through $\partial \Sigma_t$.
Thus, $J $ is conserved. 
The explicit form of $J$ is calculated as 
\begin{align}
    J&=a^3 \int^{\chi_a}_0 d\chi \int^{\pi}_0 d\theta \int^{2\pi}_0 d\phi \left( T^{(0)0\mu} +\delta T^{0\mu} \right) \xi_{\mu} \sin^2\chi \sin\theta \\
    &=\frac{(1+w)a^{3(w+1)}\bar{\rho}}{a^{3w}_{\rm max}}V_f \int^{\chi_a}_0 d\chi \int^{\pi}_0 d\theta \int^{2\pi}_0 d\phi \sin^4\chi\sin^3\theta \\
    &\simeq \frac{8(1+w)\pi a^{3(w+1)}\bar{\rho}}{15a^{3w}_{\rm max}}V_f \chi^5_a +{\mathcal O}(\chi^6_a)
\end{align}
which is indeed conserved.

\bibliographystyle{utphys}
\bibliography{draft}
	
\end{document}